\documentclass[12pt]{article}
\usepackage{textcomp}
\usepackage{graphicx,psfrag,epsf}
\usepackage{enumerate}
\usepackage{natbib}
\usepackage{url} 
\usepackage{indentfirst}
\usepackage{amssymb}
\usepackage{algorithm}
\usepackage{algorithmic}
\usepackage{hyperref}
\usepackage{booktabs}
\usepackage{amsmath,soul}
\usepackage{graphicx}
\usepackage{float}
\usepackage{xcolor}
\usepackage{comment}
\usepackage[most]{tcolorbox}
\usepackage{longtable}
\usepackage{amsthm}
\usepackage{subfig}
\usepackage{graphicx}
\providecommand{\blind}{1}
\newcommand{\argmin}{\mathop{\mathrm{arg\,min}}}

\usepackage{multirow}

\newcommand{\x}{\bm{x}}

\newcommand{\U}{\mathbf{U}}
\newcommand{\V}{\mathbf{V}}
\newcommand{\Y}{\bm{Y}}

\newcommand{\ind}{\mathbbm{1}}

\def \EE{\mathbbm{E}}
\def \PP{\mathbbm{P}}
\def \NN{\mathbbm{N}}
\def \RR{\mathbbm{R}}
\def \SS{\mathbbm{S}}
\def \OO{\mathbbm{O}}
\def \II{\mathbbm{I}}
\def \ZZ{\mathbbm{Z}}
\def \md{\mathrm{d}}

\def \Psc{\mathcal{P}}
\def \Nsc{\mathcal{N}}

\def \Ab{\mathbf{A}}

\def \X{\mathbf{X}}
\def \R{\mathbf{R}}
\def \S{\mathbf{S}}
\def \E{\mathbf{E}}
\def \D{\mathbf{D}}
\def \W{\mathbf{W}}
\def \Y{\mathbf{Y}}
\def \B{\mathbf{B}}
\def \M{\bm{M}}
\def \u{\bm{u}}
\def \v{\bm{v}}
\def \z{\bm{z}}
\def \b{\bm{b}}
\def \a{\bm{a}}
\def \c{\bm{c}}

\def\calV{{\cal  V}}
\def\calN{{\cal  N}}

\def \bmu{\boldsymbol{\mu}}

\def \bU{\mathbf{U}}
\def \bV{\mathbf{V}}
\def \bI{\mathbf{I}}
\def \bW{\boldsymbol{W}}
\def \bZ{\boldsymbol{Z}}
\def \bX{\boldsymbol{X}}
\def \bY{\boldsymbol{Y}}
\def \bA{\boldsymbol{A}}
\def \bO{\boldymbol{O}}

\def\calD{{\mathcal D}}

\def\calN{{\mathcal N}}

\def\calV{{\mathcal V}}

\def \logit{{\rm logit}}

\def\trans{^{\top}}

\def \bSigma{\boldsymbol{\Sigma}}
\def \bOmega{\boldsymbol{\Omega}}
\def \bbeta{\boldsymbol{\beta}}
\def \balpha{\boldsymbol{\alpha}}
\def \bTheta{\boldsymbol{\Theta}}

\def\wh{\widehat}
\def\wt{\widetilde}
\def\th{^{\text{th}}}

\DeclareMathOperator*{\argmax}{arg\,max}

\newtheorem{Assumption}{Assumption}

\newtheorem{Theorem}{Theorem}
\newtheorem{Corollary}{Corollary}

\newtheorem{definition}{Definition}

\def\spacingset#1{\renewcommand{\baselinestretch}%
{#1}\small\normalsize} \spacingset{1}

\def\supS{^{\scriptscriptstyle (S)}}
\def\supR{^{\scriptscriptstyle (R)}}

\usepackage{amsmath}
\usepackage{amssymb}
\usepackage{amsfonts}
\usepackage{amsthm}
\usepackage{bm}
\usepackage{bbm}
\usepackage{mathtools}
\usepackage{mathrsfs}
\usepackage{algorithm,algorithmic}
\usepackage{enumitem}

\begin{document}

\def\spacingset#1{\renewcommand{\baselinestretch}%
{#1}\small\normalsize} \spacingset{1}


\def \EE{\mathbbm{E}}
\def \PP{\mathbbm{P}}
\def \NN{\mathbbm{N}}
\def \RR{\mathbbm{R}}
\def \SS{\mathbbm{S}}
\def \OO{\mathbbm{O}}
\def \II{\mathbbm{I}}
\def \ZZ{\mathbbm{Z}}
\def \md{\mathrm{d}}

\def \Psc{\mathcal{P}}
\def \Nsc{\mathcal{N}}

\def \Ob{\mathbf{O}}
\def \Ib{\mathbf{I}}
\def \X{\mathbf{X}}
\def \R{\mathbf{R}}
\def \S{\mathbf{S}}
\def \E{\mathbf{E}}
\def \D{\mathbf{D}}
\def \W{\mathbf{W}}
\def \Y{\mathbf{Y}}
\def \B{\mathbf{B}}
\def \G{\mathbf{G}}
\def \H{\mathbf{H}}
\def \M{\bm{M}}
\def \u{\bm{u}}
\def \v{\bm{v}}
\def \z{\bm{z}}
\def \b{\bm{b}}
\def \a{\bm{a}}
\def \c{\bm{c}}
\def \g{\bm{g}}
\def \t{\bm{t}}
\def\calV{{\cal  V}} 
\def\calN{{\cal  N}} 

\def \bmu{\boldsymbol{\mu}}

\def \bU{\mathbf{U}}
\def \bV{\mathbf{V}}
\def \bI{\mathbf{I}}
\def \bW{\boldsymbol{W}}
\def \bZ{\boldsymbol{Z}}
\def \bX{\boldsymbol{X}}
\def \bY{\boldsymbol{Y}}
\def \bA{\boldsymbol{A}}
\def\bO{\boldsymbol{O}}

\def \logit{{\rm logit}}

\def \bSigma{\boldsymbol{\Sigma}}
\def \bOmega{\boldsymbol{\Omega}}
\def \bbeta{\boldsymbol{\beta}}
\def \balpha{\boldsymbol{\alpha}}
\def \bTheta{\boldsymbol{\Theta}}
\def \bdelta{\boldsymbol{\delta}}

\def\wh{\widehat}
\def\wt{\widetilde}
\def\th{^{\text{th}}}

\newcommand{\doudou}[1]{\textcolor{red}{#1}}
\newcommand{\yuming}[1]{\textcolor{orange}{#1}}
\newcommand{\yuan}[1]{\textcolor{cyan}{#1}}

\newcommand{\yzcomment}[1]{{\color{orange} \scriptsize \bf [YZ: #1]} }


\if1\blind
{
  \title{\bf Spherical Mixture Integration for Latent Embedding Alignment across Multi-Source Feature Spaces}
   \author{Yuming Zhang$^1$\thanks{Zhang and Duan contributed equally to this work.}, Congyuan Duan$^2$\footnotemark[1], Dong Xia$^2$, 
   Doudou Zhou$^3$\thanks{Zhou and Cai contributed equally as corresponding authors. Emails: \texttt{ddzhou@nus.edu.sg},  \texttt{tcai@hsph.harvard.edu}. }, Tianxi Cai$^1$\footnotemark[2] \\
   \small  $^1$Department of Biostatistics, Harvard T.H. Chan School of Public Health \\
   \small  $^2$Department of Mathematics, Hong Kong University of Science and Technology\\
    \small $^3$Department of Statistics and Data Science, National University of Singapore
    }
    \date{}
  \maketitle
} \fi

\if0\blind
{
  \bigskip
  \bigskip
  \bigskip
  \begin{center}
    {\LARGE\bf Spherical Mixture Integration for Latent Embedding Alignment across Multi-Source Feature Spaces}
\end{center}
  \medskip
} \fi
\date{}
\bigskip

\begin{abstract}
Multi-institutional electronic health record (Multi-EHR) data have emerged as a powerful resource for developing predictive models to support clinical decisions and for generating reliable real-world evidence. By aggregating information from diverse patient populations and institutions, they enhance the robustness and generalizability of models and findings. However, analyzing multi-EHR remains challenging because disparate institutions rarely map all data elements to common ontologies, and raw EHR codes are often overly granular and institution-specific, fragmenting representations of the same clinical concept. Hence, integrative analysis must overcome two key hurdles: harmonizing codes with the same clinical meaning (synonymy), and aligning institutional feature spaces. To address these challenges, we propose SMILE, a Spherical Mixture Integration for Latent Embedding alignment across multi-source feature spaces, where embeddings from heterogeneous sources serve as privacy-preserving summaries of clinical concepts and sparse relational pairs provide weak supervision. Synonymy is modeled via a mixture of von Mises-Fisher distributions, yielding unified representations of semantically equivalent raw codes. We develop a composite quasi-likelihood estimator with non-asymptotic error bounds for the latent representations and mixture mean directions and consistent synonym-cluster recovery, quantifying the gains from integrating multiple sources and knowledge-graph information. Simulations and a multi-institutional EHR application demonstrate improved alignment and synonym clustering.
\end{abstract}

\noindent%
\vfill

\newpage
\spacingset{1.8} 

\section{Introduction}
\label{sec:introduction}

\subsection{Background}

Multi-institutional electronic health record (EHR) data offer great opportunities for advancing predictive modeling to support clinical decisions and for generating high-quality real-world evidence \citep{marwaha2024mobilizing}. Jointly analyzing data across health systems enables analyses at greater scale, captures broader clinical heterogeneity, and supports findings that generalize more reliably across patient populations and care settings \citep{bianchi2024all,sittig2012survey}. Such diversity is critical for improving model transportability and mitigating site-specific bias. 

Realizing this promise, however, requires overcoming significant integration challenges. Health systems often use different coding standards, local documentation practices, and incomplete mappings to shared ontologies \citep{garcia2025multi}. Although standardized ontologies such as ICD-9 \citep{bramer1988international}, RxNorm \citep{liu2005rxnorm}, and Logical Observation Identifiers Names and Codes (LOINC) \citep{mcdonald2003loinc} provide shared vocabularies, local EHR codes are often only partially mapped to these standards \citep{gan2025arch,hong2021clinical}. Moreover, many raw EHR codes are highly granular, leading to fragmented and sparse representations of clinically similar concepts. Consequently, each institution induces its own high-dimensional feature space, and the union of these spaces across sites is heterogeneous, partially overlapping, and mis-aligned. Effective cross-site modeling therefore requires both consolidating semantically equivalent or closely related codes into coherent concept groupings and aligning heterogeneous institutional feature spaces.

Importantly, this challenge extends beyond missing data to representation-level incompatibility. Restricting analyses to features shared across sites discards valuable information, while ad hoc aggregation of unmatched codes risks introducing measurement error and distorting cross-site comparisons. Models trained on naively pooled features may also fail to generalize when clinically equivalent codes are encoded differently. These limitations motivate methods that operate directly on learned representations rather than raw codes \citep{si2021deep,choi2016multi,zhou2025representation}. 

From a statistical perspective, such multi-institutional integration can be formulated as a coupled latent variable problem. Across institutions and data modalities, observed features constitute a noisy and partially overlapping view of a shared but unobserved set of underlying clinical concepts. The inferential goals are to recover a common latent representation from these heterogeneous and incomplete views, and to identify sets of synonymous features corresponding to the same underlying concept in the absence of fully observed correspondences. Feature-space misalignment obscures latent synonymy, while failure to resolve synonymy artificially inflates dimensionality and further complicates alignment. These inter-dependencies necessitate a unified framework that jointly addresses representation alignment and latent concept recovery under partial overlap.

In practice, direct sharing of patient-level data across institutions is often infeasible due to privacy, regulatory, and governance constraints. To enable integrative analyses under these restrictions, institutions instead construct privacy-preserving summary representations of their local data that can be shared to support federated or distributed learning without exposing individual-level records \citep{li2023federated,brat2020international}. For feature alignment in particular, sites may share summary statistics that characterize feature distributions, such as co-occurrence-based measures or embedding-derived representations, to facilitate cross-institutional integration \citep{hong2021clinical,li2024multisource,zhou2025representation}. These summaries retain substantial structural information about relationships among features while avoiding direct patient-level data transfer. Accordingly, we model the observed embeddings as heterogeneous and noisy realizations of a latent geometric structure shared across institutions. The resulting statistical objective is to recover this common latent geometry together with the grouping structure induced by concept-level synonymy.

In this work, we develop \textbf{S}pherical \textbf{M}ixture \textbf{I}ntegration for \textbf{L}atent \textbf{E}mbedding (SMILE) alignment, a probabilistic framework that jointly addresses representation alignment and synonym recovery across multi-source feature spaces. SMILE integrates a multi-source latent factor model, which projects heterogeneous institutional views into a shared low-dimensional space, with a spherical mixture component that captures latent synonym groups as concentrated mean directions on the unit sphere. Auxiliary relational information from existing knowledge graphs is incorporated through likelihood-based components that propagate weak supervision across the latent geometry. By coupling alignment and clustering within a single generative framework, SMILE enables coherent recovery of harmonized representations and underlying clinical concepts rather than treating these tasks as separate sequential steps.

\subsection{Related Work}
\label{sec:related_works}

Our work is related to the multi-view integration literature, which decomposes variation into shared and view-specific components under the assumption that all views are measured on a common, aligned feature set. Representative examples include JIVE, AJIVE, and related structured factor models for multi-block data \citep{lock2013joint,feng2018angle,zhou2015group,yang2016non,gaynanova2019structural,park2020integrative,lock2022bidimensional,yi2023hierarchical}. These methods yield principled low-rank summaries and interpretable shared-versus-specific structure, but their reliance on aligned coordinates limits their applicability to our setting where each source has its own vocabulary and overlap is only partial, leaving a methodological gap for integrating views with heterogeneous feature spaces.

When feature spaces differ across sources, alignment is often formulated as learning mappings into a common latent space using geometric, distributional, or weak correspondence constraints. This regime is studied broadly under heterogeneous transfer learning and related representation-alignment formulations \citep{day2017heterogeneous,feuz2015fsr,bao2023survey}. In multi-institutional EHR studies, a growing literature instantiates this idea by aligning code representations across sites using partially shared vocabularies, ontology anchors, or privacy-preserving summary representations (e.g., embeddings or aggregated co-occurrence statistics) that can be shared without releasing patient-level records \citep{hong2021clinical,zhou2022multiview,li2023federated,brat2020international,li2024multisource,zhou2025representation}. Closely related work on embedding-space alignment (e.g., cross-lingual word embeddings) shows that meaningful alignment can be achieved with limited supervision via approximately orthogonal transformations and self-learning, while emphasizing the structural assumptions needed for identifiability \citep{alaux2018unsupervised,grave2019unsupervised}.

In multi-EHR integration, partially overlapping feature spaces are further compounded by the fact that correspondences are not merely missing but often many-to-one, in that multiple local codes may refer to the same clinical concept. This turns alignment into a joint alignment-and-clustering problem, where the goal is to recover latent synonym groups while constructing a harmonized embedding space. Existing EHR pipelines typically address this synonymy via manual ontology mappings or heuristic code aggregation, which are often incomplete across institutions and decoupled from representation alignment. Directional mixture models, such as mixtures of von Mises-Fisher distributions, provide a natural probabilistic mechanism for representing tight semantic clusters on the unit sphere \citep{gopal2014mises,barbaro2021sparse,shi2021spherical}, offering a principled alternative to such ad hoc synonym handling. Nevertheless, these methods do not tackle the joint alignment-and-clustering problem. As a result, there is currently no unified framework that jointly pools synonym and alignment evidence within a single statistical objective, especially when only weak anchors are available.

Incorporating auxiliary relational information connects our framework to the literatures on community detection, stochastic block models (SBMs), and network models with side information \citep{holland1983stochastic,lei2015consistency,hu2024network,pei2022graph}. However, our setting departs fundamentally from classic SBM regimes. In synonym recovery, the number of latent clusters grows with the vocabulary size, while most clusters remain small and may even be finite. Moreover, we must simultaneously integrate multiple heterogeneous relation types and multi-view embeddings, rather than modeling a single homogeneous network. Although SBMs with a growing number of communities have been studied \citep{choi2012stochastic,rohe2011spectral}, existing theory typically assumes community sizes also grow sufficiently fast and focuses on a single relation type. These assumptions do not cover the regime we consider, where the number of clusters diverges while cluster sizes remain small, and heterogeneous relational views must be jointly leveraged for many-to-one synonym clustering. As such, our work addresses a previously uncharacterized high-cluster-number, small-cluster-size, multi-view recovery problem that falls outside of the guarantees of existing network theory.

\subsection{Contributions} 

We propose SMILE to address the lack of a unified framework for simultaneous alignment and synonym recovery in multi-source settings characterized by heterogeneous vocabularies, partial overlap, and many-to-one correspondences. Rather than separating embedding alignment and clustering into sequential steps, SMILE treats them as a single joint estimation problem. By modeling heterogeneous embeddings and auxiliary relations as complementary views of a shared latent geometry, SMILE integrates multi-source latent factor modeling with a spherical mixture component that directly encodes concept-level grouping. Auxiliary relational information enters through logistic components that provide weak supervision, allowing the model to leverage sparse and noisy similarity signals. Estimation via a composite likelihood framework accommodates heterogeneous embedding dimensions and partial correspondence while maintaining computational tractability.

From a theoretical standpoint, we provide the first recovery guarantees tailored to the high-cluster-number, small-cluster-size, multi-view regime common in real-world EHR data integration. We establish non-asymptotic error bounds for feature-level latent representations and cluster-level mean directions, along with consistency of synonym cluster recovery as vocabulary size grows. These results clarify how estimation accuracy scales with the number and quality of sources, the degree of vocabulary overlap, and the availability of auxiliary relational information.
Empirically, we demonstrate that SMILE improves both multi-source feature alignment and synonym clustering relative to existing embedding-alignment and clustering baselines, in simulations and a real-world multi-institutional EHR application. Taken together, our results demonstrate that jointly harmonizing multiple sources, auxiliary relational information, and sparse correspondence signals can materially enhance multi-EHR integration and analysis.

\section{The SMILE Methodology}
\label{sec:method}
\subsection{Generative Model for Latent Geometry}

Our goal is to project heterogeneous, partially overlapping high-dimensional embeddings into a shared low-dimensional space to recover latent synonym groups. Let $[n]\equiv \{1,\ldots,n\}$ index distinct EHR features (both standard ontology and institution-specific codes). Embeddings are obtained from $L$ heterogeneous sources. For each source $l\in[L]$, let $\mathcal{S}_l\subset[n]$ be the subset of features with embeddings from that source, with size $n_l\equiv |\mathcal{S}_l|$. We assume $\cup_{l=1}^L \mathcal{S}_l = [n]$, so every feature appears in at least one source, though overlaps may be partial. 
Define $N\equiv \sum_{l=1}^L n_l$ as the total number of feature instances across sources; the ratio $N/n$ is the average number of source-level replicates per feature. For each feature $i\in\mathcal{S}_l$, let $\mathbf{U}_i^{(l)}\in\mathbb{R}^{r_l}$ be its unit-norm embedding from source $l$, where $r_l$ may be large. In practice, these source-specific embeddings are typically learned from site-specific co-occurrence patterns \citep{arora2015latent} or from pretrained language models (PLMs) applied to semantic code descriptions (e.g., \citealt{liu2020self,yuan2022coder}). To achieve alignment and clustering, our framework SMILE specifies a joint generative model with four components: a multi-source factor model for the embeddings, a spherical mixture for synonym groups, and logistic models for auxiliary relational information, as detailed below.

First, we model the high-dimensional, source-specific embeddings using a multi-source latent factor model. Each feature $i$ has a shared, unit-norm latent representation $\V_i\in\RR^r$, where $r\ll r_l$ and $r\geq 2$. Conditional on $\{\V_i\}_{i\in[n]}$, the observed embeddings follow a multi-view latent factor model
\begin{equation}
\label{eqn:latent_factor_model}
    \U_i^{(l)} = \W_l \V_i + \E_i^{(l)}, \qquad i\in\mathcal{S}_l,
\end{equation}
where $\W_l\in\RR^{r_l\times r}$ is a source-specific loading matrix of rank $r$, and the noise vector $\E_i^{(l)}$ has independent mean-zero sub-Gaussian entries with variance $\sigma^2$. Projection into this shared latent space reconciles the partially overlapping feature sets across sources, with $\{\V_i\}_{i \in [n]}$ identifiable only up to a global orthogonal transformation. For brevity, we henceforth write $\{\V\}$ for the collection of $\{\V_i\}_{i \in [n]}$.

Second, to recover the latent synonym groups, we assume that the shared representations $\{\V\}$ arise from a mixture of $K$ von Mises-Fisher (vMF) distributions on the unit hypersphere:
\begin{equation}
\label{eqn:vmf_model}
    \V_i \sim f_r(\x;\bmu_{z_i},\kappa), \qquad
    f_r(\x;\bmu,\kappa)=C_r(\kappa)\exp(\kappa\bmu^\top\x),
\end{equation}
where $z_i\in[K]$ denotes the latent cluster membership (synonym group) of feature $i$, $\bmu_k$ is the unit-norm mean direction of cluster $k$, and $\kappa\ge0$ is a common concentration parameter. The normalizing constant is given by $C_r(\kappa)=\kappa^{r / 2-1} /\{(2 \pi)^{r / 2} I_{r / 2-1}(\kappa)\}$, with $I_v(\cdot)$ denoting the modified Bessel function of the first kind at order $v$. By modeling each mean direction $\bmu_k$ as a concept-level embedding, this formulation allows synonymous features to realistically fluctuate around a shared center rather than collapsing to a single point, capturing source-specific variation while preserving the geometric cohesion of the groups. In practice, the total number of clusters $K$ can be selected in a data-driven manner, typically guided by auxiliary domain knowledge that provides partial cluster membership labels, such as existing medical ontologies in the specific case of EHR data.

Finally, we incorporate sparsely observed auxiliary knowledge graph information through two logistic models to supervise the latent geometry. To enforce broad concept-level coherence, the first model links similarity labels $\delta_{ij}\supS\in\{0,1\}$ (e.g., ``Acetaminophen'' and ``Ibuprofen'') to the cluster means:
\begin{equation}
\label{eqn:similarity_model}
    \PP(\delta_{ij}\supS=1 | \bmu_{z_i},\bmu_{z_j},\beta_1,\beta_2)    =
    \frac{\exp(\beta_1+\beta_2\bmu_{z_i}^\top\bmu_{z_j})}
         {1+\exp(\beta_1+\beta_2\bmu_{z_i}^\top\bmu_{z_j})},
\end{equation}
where $\beta_1$ controls observation sparsity and $\beta_2$ dictates geometric alignment. To refine fine-grained, feature-level geometry, the second model links broader functional relatedness labels $\delta_{ij}\supR\in\{0,1\}$ (e.g.,  ``Type 2 Diabetes'' and ``Metformin'') directly to the latent representations:
\begin{equation}\label{eqn:relatedness_model}
    \PP(\delta_{ij}\supR=1 | \V_i,\V_j,\beta_3,\R)    =
    \frac{\exp(\beta_3+\V_i^\top\R\V_j)}
         {1+\exp(\beta_3+\V_i^\top\R\V_j)},
\end{equation}
where $\beta_3$ controls observation density and the symmetric matrix $\R$ weights the latent dimensions. Crucially, neither relation requires fully observed pairs across the feature space.

\subsection{Composite Quasi-likelihood Estimation}

To estimate the latent embeddings $\{\V\}$, concept-level means $\{\bmu\}$, and cluster labels $\{z\}$, we adopt a composite quasi-likelihood approach. Because the observed data combine multi-source embeddings with sparsely sampled relational pairs, a fully specified joint likelihood is neither natural nor necessary. Instead, we construct a unified, computationally tractable objective by summing the conditional (quasi-)likelihoods of the four model components \eqref{eqn:latent_factor_model}--\eqref{eqn:relatedness_model}, estimating all other parameters jointly as nuisance variables.

For the multi-source latent factor model \eqref{eqn:latent_factor_model}, we enforce the geometric alignment of heterogeneous embeddings using a dimension-normalized squared loss
\begin{equation}
    \ell_{\rm lr}\big( \{\V\}, \{\W\} \big)
    = \frac{1}{N} \sum_{l=1}^L \sum_{i\in\mathcal{S}_l}
    \frac{1}{r_l} \left\|\U_i^{(l)} - \W_l \V_i\right\|_2^2 .
\end{equation}
For the vMF mixture \eqref{eqn:vmf_model}, we enforce the concentration of latent embeddings around their concept-level mean directions by minimizing the negative log-likelihood:
\begin{equation}
\ell_{\rm vmf}\big( \{\V\}, \{\bmu\}, \kappa \big) = 
-\frac{1}{n\kappa}\sum_{i=1}^n  \log \left\{ \sum_{k=1}^K f_r({\V}_i; \bmu_k, \kappa) \pi_{ik} \right\}, \quad \mbox{with } \pi_{ik}\equiv \PP(z_i=k).
\end{equation}
Here, the prior probabilities $\{\pi_{ik}\}$ allow us to flexibly incorporate partial knowledge of cluster membership (e.g., from existing ontologies or feature-type constraints) as probabilistic guidance for synonym recovery, without imposing hard labels.

Finally, we formulate the negative log-likelihoods for the sparsely observed relational data based on their respective positive ($\mathcal{P}_i$) and negative ($\mathcal{N}_i$) sets, $\Omega_i\supS \equiv \mathcal{P}_i\supS \cup \mathcal{N}_i\supS$ and $\Omega_i\supR \equiv \mathcal{P}_i\supR \cup \mathcal{N}_i\supR$, with total observed pair counts denoted by $n_S $ and $n_R$. The loss for similarity pairs \eqref{eqn:similarity_model} links annotations directly to concept-level geometry:
\begin{equation}
\begin{aligned}
    \ell_{\rm sim} \big(\{\bmu\}, \beta_1, \beta_2 \big) =& -\frac{1}{n_S} \sum_{i=1}^n \sum_{j\in \Omega_i\supS} \log \left\{ \sum_{k_1, k_2 \in [K]} \PP(\delta_{ij}\supS | \bmu_{k_1},\bmu_{k_2},\beta_1,\beta_2) \pi_{ik_1} \pi_{jk_2}  \right\},
\end{aligned}
\end{equation}
and the loss for related pairs \eqref{eqn:relatedness_model} informs the fine-scale latent geometry independently of cluster membership:
\begin{equation}
    \ell_{\rm rel}\big( \{ \V \}, \beta_3, \R \big) = -\frac{1}{n_R} \sum_{i=1}^n \sum_{j\in \Omega_i\supR} \log \PP(\delta_{ij}\supR | \V_i, \V_j,\beta_3,\R), 
\end{equation}
where $\PP(\delta_{ij}\supS | \bmu_{k_1},\bmu_{k_2},\beta_1,\beta_2)$ and $\PP(\delta_{ij}\supR | \V_i, \V_j,\beta_3,\R)$ are the standard Bernoulli likelihoods derived from the logistic formulations in \eqref{eqn:similarity_model} and \eqref{eqn:relatedness_model} respectively.

Bringing these components together, the composite estimator is defined as the minimizer of the weighted sum of the four losses:
\begin{equation}
\begin{aligned}
\label{def:composite_objective}
    \{\wh{\V}\}, \{\wh{\W}\}, \{\wh{\bmu}\}, \wh{\kappa}, \wh{\beta}_1, \wh{\beta}_2, \wh{\beta}_3, \wh{\R} \equiv   & \argmin\;   \ell_{\rm lr}\big( \{\V\}, \{\W\} \big) + w_{\rm vmf} \; \ell_{\rm vmf}\big( \{\V\}, \{\bmu\}, \kappa \big) + \\
    & \quad w_{\rm sim}\; \ell_{\rm sim} \big(\{\bmu\}, \beta_1, \beta_2 \big) + w_{\rm rel}\; \ell_{\rm rel} \big(\{\V\}, \beta_3, \R \big),
\end{aligned}
\end{equation}
subject to the geometric constraints $\|\V_i\|_2=1$ for all $i$, $\|\bmu_k\|_2=1$ for all $k$, $\kappa\geq 0$, and symmetric $\R$. The weights $w_{\rm vmf}, w_{\rm sim}, w_{\rm rel} \geq 0$
control the relative influence of each model component. Since each loss is
normalized by its effective sample size, these weights primarily serve to balance heterogeneous noise levels and varying information content across the different data sources. In practice, they are selected in a data-driven manner and yield stable performance under moderate variations.

The estimated cluster of each feature is assigned based on the vMF mixture posterior: given the fitted $\{\wh{\V}\},\{\wh{\bmu}\},\wh\kappa$,
\begin{equation}\label{eqn:cluster_assignment}
    \wh{z}_i\equiv\argmax_{k\in[K]}\gamma_{ik},\quad \text{with } \gamma_{ik}\equiv\PP(z_i=k\mid\wh{\V}_i)=\frac{f_r(\wh{\V}_i;\wh{\bmu}_k,\wh\kappa)\,\pi_{ik}}{\sum_{k'=1}^K f_r(\wh{\V}_i;\wh{\bmu}_{k'},\wh\kappa)\,\pi_{ik'}}.
\end{equation}

\subsection{Algorithm}

The composite objective \eqref{def:composite_objective} is optimized via the alternating blockwise procedure in Algorithm~\ref{algo:SMILE}. We first initialize the feature-level embeddings $\{\V\}$ by fitting only the multi-source latent factor and relatedness components, avoiding concept-level structure and thus requiring neither cluster labels nor mean directions. We then initialize the concept-level parameters. In practice, careful initialization of the mean directions $\{\bmu\}$ is important and can be obtained, for example, by applying spherical $k$-means to the initial embeddings. Because the synonym clusters are latent, we use  an expectation-maximization (EM) step to update the vMF mixture parameters and cluster assignments given the current embeddings, and alternate this with gradient-based updates of the embeddings and feature-level parameters under the relevant geometric constraints. Across iterations, the procedure integrates synonymy encoded by the vMF mixture and auxiliary relational pairs within a single composite objective.

\begin{algorithm}[!tb]
\caption{The SMILE algorithm}
\label{algo:SMILE}
\begin{algorithmic}[1]
\renewcommand{\algorithmicrequire}{\textbf{Input:}}
\renewcommand{\algorithmicensure}{\textbf{Output:}}
\REQUIRE Source-specific embeddings $\{\U\}$, relational pairs $\{\delta\supS\}, \{\delta\supR\}$, prior probabilities $\{\pi_{ik}\}$, weights $\{w_{\rm vmf}, w_{\rm sim}, w_{\rm rel}\}$, initial values for all parameters.   
\ENSURE Estimates for $\{\V\}, \{\bmu\}, \{z\}$.
\\ \textbf{Step 1 (Initialization):} Set $s = 1$. Initialize $\{\V\}$ by minimizing \eqref{def:composite_objective} with $w_{\rm vmf} = w_{\rm sim} = 0$, yielding $\{\wh{\V}^{(s-1)}\}, \{\wh{\W}^{(s-1)}\}, \wh{\beta}_3^{(s-1)}, \wh{\R}^{(s-1)}$. This step uses only feature-level alignment and relations, requiring no cluster labels or mean directions.
\\ \textbf{Step 2 (Concept-level update):} Fixing $\{\wh{\V}^{(s-1)}\}$, update $\{\bmu\}, \kappa, \beta_1, \beta_2, \{z\}$ by minimizing \eqref{def:composite_objective} with $w_{\rm rel} = 0$ using an EM algorithm (see Supplement~A for details), yielding 
$\{\wh{\bmu}^{(s)}\}, \wh{\kappa}^{(s)}, \wh{\beta}_1^{(s)}, \wh{\beta}_2^{(s)}, \{\wh{z}^{(s)}\}$. This step updates concept-level geometry and cluster assignments given fixed feature-level embeddings.
\\ \textbf{Step 3 (Feature-level refinement):} Fixing $\{\wh{\bmu}^{(s)}\}, \wh{\kappa}^{(s)}, \{\wh{z}^{(s)}\}$, refine $\{\V\}, \{\W\}, \beta_3, \R$ by minimizing \eqref{def:composite_objective} with $w_{\rm sim} = 0$, yielding $\{\wh{\V}^{(s)}\}, \{\wh{\W}^{(s)}\}, \wh{\beta}_3^{(s)}, \wh{\R}^{(s)}$. This step uses only terms involving feature-level variables.
\\ \textbf{Step 4 (Iteration):} Set $s \leftarrow s+1$ and repeat Steps 2--3 until convergence.
\end{algorithmic} 
\end{algorithm}

\section{Theoretical Properties}
\label{sec:theory}

This section establishes the identifiability of the SMILE model (Section~\ref{sec:identifiability}) and theoretical guarantees for the SMILE estimator (Sections~\ref{sec:assumptions} and \ref{sec:theorectical_results}). We focus on three theoretical results: a non-asymptotic recovery bound for the latent embeddings $\{\V\}$, a corresponding bound for the cluster mean directions $\{\bmu\}$, and consistency of the induced clustering $\{\wh{z}\}$.  
Throughout, for a matrix $\Ab$, let $\|\Ab\|$ and $\|\Ab\|_F$ denote its operator (spectral) and Frobenius norms, respectively. For a vector $\a$, let $\|\a\|_2$ denote the Euclidean norm. Define $\mathbb{O}^{r\times r}\equiv \{\Ob\in\RR^{r\times r}:\Ob^\top\Ob=\Ib_r\}$ as the orthogonal group. We write $a\lesssim b$ and $b\gtrsim a$ if $a\le Cb$ for some finite positive constant $C$, and $a\wedge b\equiv \min\{a,b\}$. The indicator function is denoted by $\ind(\cdot)$.

\subsection{Identifiability of the SMILE Model}
\label{sec:identifiability}

The SMILE model is identifiable only up to two natural transformations of its latent parameters: a common rotation of the latent space and a permutation of cluster labels that preserves the prior probabilities $\{\pi_{ik}\}$. Specifically, for any $\Ob \in \mathbb{O}^{r\times r}$ and any permutation $\tau$ of $[K]$ satisfying $\pi_{i,\tau(k)} = \pi_{ik}$ for all $i\in[n], k\in[K]$, 
the transformation
\begin{equation*}
    \V'_i = \Ob \V_i, \qquad \bmu'_k = \Ob \bmu_{\tau(k)}, \qquad z'_i = \tau^{-1}(z_i), \qquad \W'_l = \W_l \Ob^\top, \qquad \R' = \Ob \R \Ob^\top,
\end{equation*}
leaves the joint distribution of the observed quantities $\{\U\}, \{\delta\supS\}, \{\delta\supR\}$ unchanged. Indeed, the multi-view factor model in \eqref{eqn:latent_factor_model} depends on the latent parameters only through $\W'_l \V'_i = \W_l \V_i$, while the relatedness model in \eqref{eqn:relatedness_model} satisfies $(\V'_i)^\top \R' \V'_j = \V_i^\top \R \V_j$. Likewise, the similarity model in \eqref{eqn:similarity_model} depends on the cluster mean only through inner products, which are preserved by a common rotation and the relabeling $z'_i = \tau^{-1}(z_i)$. The cluster prior is also unchanged because $\tau$ preserves $\{\pi_{ik}\}$.

Thus, throughout this section, $\V_i, \bmu_k, z_i, \W_l, \R$ denote one fixed representative of the true equivalence class. The orthogonal matrix $\Ob$ and permutation $\tau$ appearing in the theorem statements below serve only to align this representative with the SMILE estimator (numerically, the algorithm's output whose coordinate frame is set by the non-convex optimization).

\subsection{Assumptions}
\label{sec:assumptions}

The analysis below follows the well-initialized non-convex $M$-estimator framework developed for the EM algorithm and regularized non-convex $M$-estimators \cite{balakrishnan2017statistical,loh2015regularized}; see also \cite{wang2015high, ma2020implicit}. As in that framework, we study the theoretical guarantees when the SMILE estimator lies in a neighborhood of the truth (Definition~\ref{def:neighborhood}), which we state as Assumption~\ref{assump:neighborhood} and discuss below.

\begin{definition}[Neighborhood of the truth]
\label{def:neighborhood}
The parameter values $\{\V'\}$, $\{\bmu'\}$, $\beta_1', \beta_2', \beta_3', \R', \kappa'$ lie in a neighborhood of the truth if $\max_{i\in[n]} \|\V_i' - \V_i\|_2$, $\max_{k\in[K]} \|\bmu_k' - \bmu_k\|_2$, $|\beta_j' - \beta_j|$ for $j=1,2,3$, $\|\R' - \R\|$, and $|\kappa' - \kappa|$ are each bounded above by sufficiently small positive constants.
\end{definition}

\begin{Assumption}[Local solution]
\label{assump:neighborhood}
There exists an orthogonal alignment $\Ob\in\mathbb{O}^{r\times r}$ under which the SMILE estimator lies in the neighborhood of the truth (Definition~\ref{def:neighborhood}).
\end{Assumption}

Assumption~\ref{assump:neighborhood} is met in our setting through two key ingredients. First, Algorithm~\ref{algo:SMILE} is well initialized near the truth: the low-rank multi-source factor model makes the matrix-factorization initialization of $\{\V\}$ accurate, spherical $k$-means on the resulting embeddings yields an accurate initialization of the mean directions $\{\bmu\}$, and the anchor codes with known cluster membership, encoded through the priors $\{\pi_{ik}\}$, initialize the labels $\{z\}$ accurately. Second, the composite loss is locally well-behaved: the local strong convexity and Hessian regularity of Assumption~8 make the truth the unique nearby stationary point, toward which the blockwise updates contract, so an estimator started near the truth remains there. Therefore, Assumption~\ref{assump:neighborhood} is a reasonable hypothesis, and it suffices to develop our theoretical guarantees for this local solution.

\begin{Assumption}[Number of relational pairs]
\label{assump:samplesize}
(a) The available relatedness pairs satisfy $n_R \gtrsim n\log n$;  (b) the available similarity pairs satisfy $n_S \gtrsim Kr^2\log(rK)$.
\end{Assumption}

Part~(a) requires each feature to have $\gtrsim\log n$ related pairs on average. Part~(b), together with cluster-size balance, requires each cluster to have $\gtrsim r^2\log(rK)$ similarity pairs for estimating its mean direction; thus both conditions require only near-logarithmic average degree. We also impose standard local regularity conditions for well-initialized non-convex $M$-estimators \citep{balakrishnan2017statistical,loh2015regularized}, including a minimum cluster size, non-degeneracy and identifiability of model components, cluster separability, bounded logistic probabilities, and local Hessian regularity (Assumptions~3--8). Formal statements and discussion are deferred to Supplement~D.

\subsection{Theoretical Guarantees}
\label{sec:theorectical_results}

We now state the main theoretical guarantees for the SMILE estimator. The results quantify recovery of the latent embeddings ${\V}$, recovery of the cluster mean directions ${\bmu}$, and consistency of the induced clustering ${\widehat z}$. The bounds reflect the three sources of information in the SMILE objective: multi-view linear observations of ${\V}$, relational observations from the logistic models, and spherical mixture structure for the
latent clusters.

\begin{Theorem}[Recovery of latent embeddings]
\label{thm:estimationV-simple}
    Suppose Assumptions~\ref{assump:neighborhood}--8 hold, and let $\Ob$ be the orthogonal alignment of Assumption~\ref{assump:neighborhood}. Then, there exist choices of the weights $w_{\rm rel}$ and $w_{\rm vmf}$ such that, with probability at least $1-9(nL)^{-50}-3(rK)^{-49}$, we have
    \begin{equation*}
        \frac{1}{n}\sum_{i=1}^n \|\widehat{\V}_i- \Ob \V_i\|_2^2\lesssim \frac{\log(nL)}{n}\left(\frac{C_1\sigma r N^{-1/2}  + C_2w_{\rm rel} n_R^{-1/2} }{D_0N^{-1}+ C_3w_{\rm rel}(rn)^{-1}}\right)^2  \wedge \left(\frac{r-1}{\kappa} + \frac{Kr^2\log(rK)}{n_S}\right),
    \end{equation*}
    where $D_0\equiv \min_{i\in[n]} |\calD_i|$ with $\mathcal{D}_i\equiv \{l\in[L]: i\in\mathcal{S}_l\}$, and $C_1,C_2, C_3$ are finite positive constants.
\end{Theorem}

The bound contains contributions from the multi-view embeddings through the source overlap $D_0$, the relatedness pairs through $n_R$, and the spherical mixture through $\kappa$ and the similarity supervision $n_S$. In particular, $\{\wh\V\}$ is consistent when $D_0^2 n\gg r^2 N\log(nL)$ (multi-view), $n_R\gg r^2n\log(nL)$ (relatedness), or $\kappa\to\infty$ with $n_S\gg Kr^2\log(rK)$ (spherical). Corollary~\ref{cor:simplified-rate-V} below specializes the bound to make these contributions explicit.

\begin{Corollary}
\label{cor:simplified-rate-V}
Under the same conditions as Theorem~\ref{thm:estimationV-simple}, with probability at least $1-9(nL)^{-50}-3(rK)^{-49}$, we have
\begin{equation*}
\frac{1}{n}\sum_{i=1}^n \big\|\widehat{\V}_i-\Ob\V_i\big\|_2^2
\;\lesssim\;
\left(\frac{r^2N\log(nL)}{D_0^2n}\right)
\;\wedge\;
\left(\frac{r^2 n\log(nL)}{n_R}\right)
\;\wedge\;
\left(\frac{r-1}{\kappa} + \frac{Kr^2\log(rK)}{n_S}\right).
\end{equation*}
\end{Corollary}

The first term is the multi-view baseline obtained when $w_{\rm rel}=w_{\rm vmf}=0$, corresponding to estimation of $\{\V\}$ from the embeddings $\{\U\}$ alone; its dependence on $D_0$ shows that higher source overlap increases effective sample size. The second term, attained as $w_{\rm rel}\to\infty$, is the relatedness-only rate, which improves on the multi-view baseline once $n_R \gtrsim D_0^2 n^2/N$; in the fully-overlapping case, where $N=Ln,D_0=L$, this reduces to $n_R\gtrsim Ln$, so even sparse relational supervision can sharpen the multi-view estimate.
The final term, attained as $w_{\rm vmf}\to\infty$, is the spherical-prior contribution. The vMF component does not localize $\{\V\}$ on its own, but ties each $\V_i$ to its mean $\bmu_{z_i}$, transferring the mean estimation accuracy from Theorem~\ref{thm:estimation-mu}, up to the within-cluster dispersion $(r-1)/\kappa$. Increasing $\kappa$ therefore improves this term only until the similarity resolution limit $Kr^2\log(rK)/n_S$ is reached.
We next control the estimation of the cluster mean directions $\{\bmu\}$.

\begin{Theorem}[Recovery of cluster mean directions]
    \label{thm:estimation-mu}
    Suppose Assumptions~\ref{assump:neighborhood}--8 hold, and let $\Ob$ be the orthogonal alignment of Assumption~\ref{assump:neighborhood}. Then, with probability at least $1-9(nL)^{-50}-3(rK)^{-49}$,
    \begin{align*}
        \frac{1}{K}\sum_{k=1}^{K} \|\widehat{\bmu}_k-\Ob\bmu_k\|_2^2\lesssim \left(\frac{Kr^2\log(rK)}{n_S}\right) \wedge \left(\frac{K r\log(rK) }{n\kappa}\;+\;\underline{E}_V^2\right),
    \end{align*}
    where $\underline{E}_V^2\equiv\tfrac{\log(nL)}{n}\big(\tfrac{C_1\sigma r N^{-1/2}+C_2 w_{\rm rel} n_R^{-1/2}}{D_0 N^{-1}+C_3 w_{\rm rel}(rn)^{-1}}\big)^2$ is the multi-view--relatedness recovery rate for $\{\V\}$ from Theorem~\ref{thm:estimationV-simple}.
\end{Theorem}

Theorem~\ref{thm:estimation-mu} gives two routes for recovering the mean directions $\{\bmu\}$. The first uses similarity pairs directly and improves as $n_S$ grows. The second uses the spherical mixture structure: each mean direction is estimated from its assigned embeddings, so its rate is controlled by the embedding accuracy $\underline{E}_V^2$ plus the within-cluster dispersion $Kr\log(rK)/(n\kappa)$. Increasing $\kappa$ reduces this dispersion until the embedding resolution limit is reached. Thus, $\{\wh\bmu\}$ is consistent if either $n_S\gg Kr^2\log(rK)$, or $n\kappa\gg Kr\log(rK)$ together with consistent recovery of $\{\wh{\V}\}$.

We next establish clustering consistency using a margin-based argument from classification and mixture model theory (see e.g., \citealt{tsybakov2004optimal, fromont2006functional}). Define the population margin for feature $i$ by
\begin{equation}\label{eq:margin}
    m_i \equiv \log \pi_{iz_i} + \kappa \bmu_{z_i}^{\top}\V_i-\max_{k\neq z_i} ( \log \pi_{ik} + \kappa \bmu_k^{\top}\V_i).
\end{equation}
This quantity measures separation between the true cluster and its closest competitor. Since errors occur only near the decision boundary or when the latent quantities are poorly estimated, the clustering error is controlled by the fraction of small-margin features plus the embedding and mean estimation errors.

\begin{Theorem}[Clustering consistency]
\label{thm:clustering}
Let $E_V^2$ and $E_\mu^2$ be upper bounds for $\frac{1}{n}\sum_{i=1}^n \|\wh{\V}_i - \Ob\V_i\|_2^2$ and $\frac{1}{K}\sum_{k=1}^K \|\wh{\bmu}_k - \Ob\bmu_k\|_2^2$, respectively, as provided by Theorems~\ref{thm:estimationV-simple} and \ref{thm:estimation-mu}. Suppose $E_V^2=o(1)$ and $E_\mu^2=o(1)$. Then, under the assumptions of those theorems, there exists a permutation $\tau$ of $[K]$ such that, with probability at least $1-9(nL)^{-50}-3(rK)^{-49}$, the misclassification rate obeys
\begin{equation*}
    \frac{1}{n} \sum_{i=1}^n \ind\{\tau(\wh{z}_i)\neq z_i\} \;\lesssim\; \frac{1}{n}\big|\{i\in[n]: m_i\leq c_m\kappa\}\big| \;+\; E_V^2 \;+\; E_\mu^2,
\end{equation*}
where $c_m$ is a finite positive constant given in Assumption~6. In particular, the right-hand side tends to $0$, so $\frac{1}{n} \sum_{i=1}^n \ind\{\tau(\wh{z}_i)= z_i\} \rightarrow 1$.
\end{Theorem}

The clustering error has two sources: estimation error $E_V^2+E_\mu^2$, and the fraction of features with small population margin. The first vanishes when $\{\V\}$ and $\{\bmu\}$ are recovered consistently, and the second vanishes under the cluster-separation condition imposed in Assumption~6. SMILE therefore recovers the latent synonym groups consistently, with sharper embeddings and means translating into sharper clustering.

\section{Simulation Studies}
\label{sec:simulations}

We have conducted extensive simulation studies to evaluate the finite-sample performance of SMILE. We specifically examine how estimation and clustering performance depend on the number of embedding sources $L$, the spherical concentration parameter $\kappa$, the number of clusters $K$, and the amount of relational supervision. We compare SMILE against three multi-view integration baselines: SVD applied to concatenated embeddings, AJIVE (Angle-based Joint and Individual Variation Explained; \citealt{feng2018angle}), and SLIDE (Structural Learning and Integrative DEcomposition of multi-view data; \citealt{gaynanova2019structural}). Since these methods only estimate embeddings and cannot directly infer cluster membership, we pair each with $k$-means and hierarchical clustering to yield six competing procedures. Cluster centers are then computed using normalized within-cluster averages, followed by permutation alignment. All reported results are based on $100$ Monte Carlo replications for each configuration.

Throughout, we consider $n=1000$ unique features and assign them to $K \in\{50, 100\}$ clusters via a uniform multinomial distribution, resulting in probabilistically balanced cluster sizes. The elements of the cluster centers $\{\bmu\}$ in the $r = 6$ dimensional latent space are drawn independently from $\mathcal{N}(1/\sqrt{6}, 1)$ and then each vector $\bmu_k$ is normalized to have unit norm. Conditional on $\{\bmu\}$ and $\{z\}$, the latent embeddings $\{\V\}$ are generated from vMF distributions with concentration parameter $\kappa\in\{100,150,200\}$. We consider $L\in\{2,3,4\}$ sources, with feature sets $\mathcal{S}_1 = \{1,\ldots,1000\}$, $\mathcal{S}_2 = \{1,\ldots,500\}$, $\mathcal{S}_3 = \{301,\ldots,800\}$, and $\mathcal{S}_4 = \{501,\ldots,1000\}$. Moreover, all source-specific embeddings $\{\U\}$ have dimension $r_l = 200$, and are generated from vMF distributions with mean direction $\W_l\V_i / \|\W_l\V_i\|_2$ and concentration parameter $60$, where $\W_l$ has entries drawn independently from $\mathcal{N}(0.6,1)$.

Because our sources only partially overlap but the competing methods require complete embedding matrices, we impute the missing embeddings. Since $\mathcal{S}_1$ contains all features, we perform SVD on $\{\U_i^{(1)}\}$ and then map the resulting embeddings into the space of source $l\in\{2,3,4\}$ using an orthogonal Procrustes transformation estimated from the observed $\{\mathbf{U}_i^{(l)}\}$. In contrast, SMILE naturally accommodates partially overlapping feature spaces and does not require imputation. 

For SMILE, we additionally generated similarity and relatedness edges according to the logistic models \eqref{eqn:similarity_model} and \eqref{eqn:relatedness_model} with $\beta_1=-0.125$, $\beta_2=5$, $\beta_3=-0.125$, and $\R = \mbox{diag}(5,\ldots,5)$. Among all positive pairs, we retain only a random subset as observed. For negative pairs, we select ``hard negatives'' with the largest cosine similarities (based on first-source embeddings). We vary the overall proportion of observed relational pairs as follows: (i) 4\% similarity and 4\% relatedness (both with 1.5\% positive, 2.5\% negative); (ii) 6\% similarity and 6\% relatedness (both with 2.5\% positive, 3.5\% negative); and (iii) 8\% similarity and 8\% relatedness (both with 3.5\% positive, 4.5\% negative).

SMILE also incorporates prior correspondence information through $\{\pi_{ik}\}$. To emulate such information, we treat 70\% of the clusters as anchor clusters, containing approximately 70\% of all features with known memberships. We transform these anchor relations into soft cluster-assignment probabilities using a spectral-clustering-based procedure (see Supplement~A for details). Clustering accuracy for all methods is evaluated on the remaining 30\% of features in the 30\% non-anchor clusters.

We considered the following settings for comparison:
\begin{itemize}
    \item \textbf{Setting 1: Varying the number of embedding sources $L$.} We evaluate each method for $L \in \{2,3,4\}$, using feature sets $\{\mathcal{S}_l : l \in [L]\}$, while fixing $\kappa = 150$, $K = 50$, and 6\% observed relational pairs for SMILE.
    \item \textbf{Setting 2: Varying the concentration parameter $\kappa$ in the vMF distributions.} We consider $\kappa \in \{100,150,200\}$, while fixing $L = 3$ with feature sets $\{\mathcal{S}_l : l \in [L]\}$, $K = 50$, and 6\% observed relational pairs for SMILE.
    \item \textbf{Setting 3: Varying the number of clusters $K$.} We consider $K \in \{50, 100\}$, while fixing $L = 3$ with feature sets $\{\mathcal{S}_l : l \in [L]\}$, $\kappa=150$, and 6\% observed relational pairs for SMILE.
    \item \textbf{Setting 4: Varying the amount of relational supervision used by SMILE.} We consider $\{4\%, 6\%, 8\%\}$ observed relational pairs for SMILE, while fixing $L = 3$ with feature sets $\{\mathcal{S}_l : l \in [L]\}$, $\kappa=150$, and $K=50$. 
\end{itemize}

To assess how well each method recovers the true latent embeddings (up to rotation), we define an accuracy metric 
\begin{equation*}
    \text{RelAcc}(\V) = \frac{1}{1+\text{RelErr}(\V)} \quad \text{with} \quad \text{RelErr}(\V) = \frac{\|\wh{\V}\wh{\V}\trans-\V_0\V_0\trans\|_F} {\|\V_0\V_0\trans\|_F },
\end{equation*}
where $\wh{\V}$ and $\V_0$ are the $n\times r$ estimated and true embedding matrices. Cluster-mean accuracy is defined analogously, using permutation-aligned estimated clusters. Clustering accuracy is evaluated by adjusted mutual information (AMI), which adjusts for chance agreement and ranges from 0 to 1.

\begin{figure}[!tbp]
    \centering
    \includegraphics[width=\textwidth]{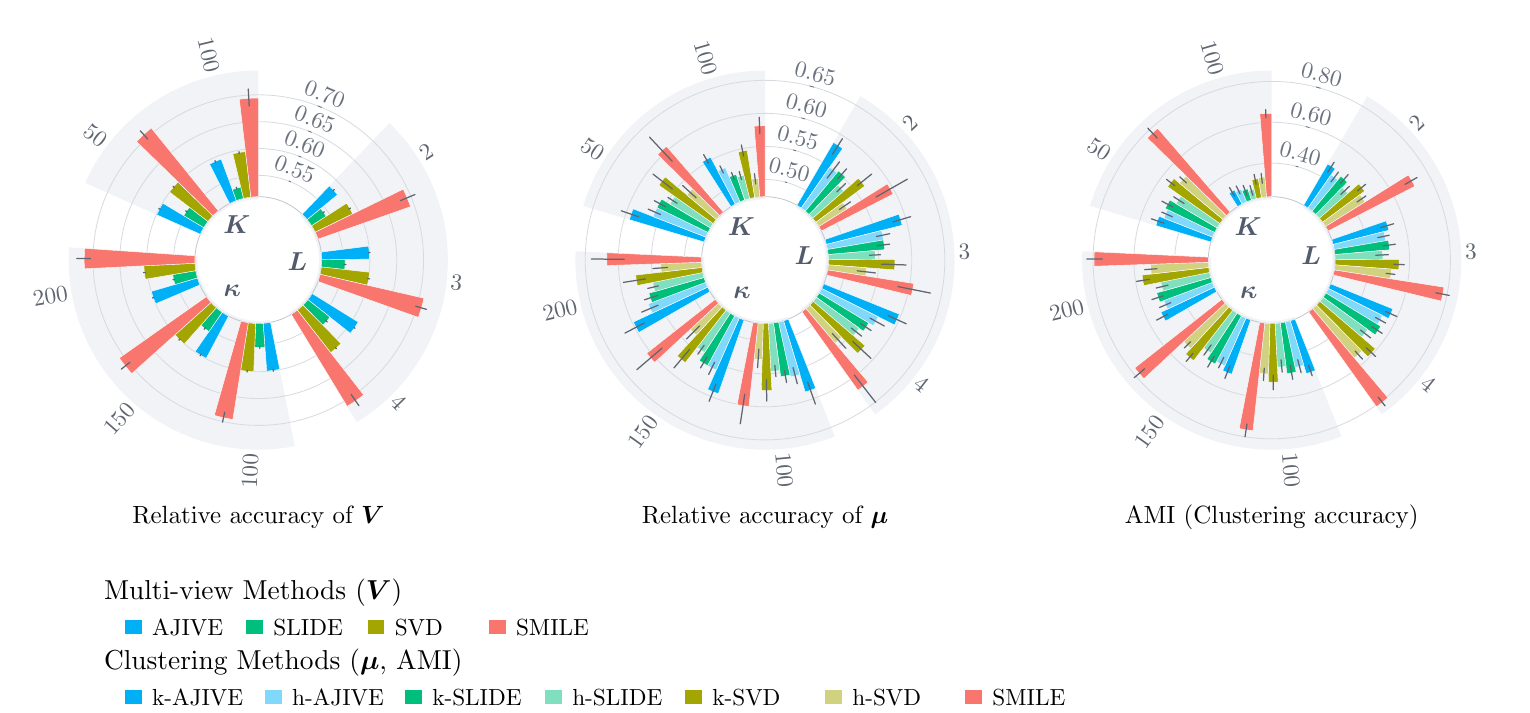}
    \caption{Empirical performance of multi-view integration and clustering methods across three simulation settings, shown as circular bar charts. Panels report relative accuracy of the latent embeddings $\{\V\}$ (left), relative accuracy of cluster centers $\{\bmu\}$ (middle), and clustering accuracy measured by AMI (right). Within each panel, the three shaded sections correspond to Setting~1: varying the number of embedding sources $L$; Setting~2: varying the vMF concentration $\kappa$; and Setting~3: varying the number of clusters $K$. Each bar gives the median estimate over the 100 replications, and the radial tick marks the interquartile range (25th--75th percentile) across replications. ``k-XXX'' and ``h-XXX'' refer to method XXX (SVD, AJIVE, or SLIDE) combined with $k$-means and hierarchical clustering, respectively.}
    \label{fig:simu_combined}
\end{figure}

Figure~\ref{fig:simu_combined} summarizes the estimation and clustering performance of all methods in Settings~1--3. The main conclusion is that SMILE consistently achieves the best performance across all settings and tasks, highlighting the benefit of jointly leveraging relational supervision and prior information to refine embeddings and recover cluster means and memberships. More specifically, for the estimation of the latent embeddings $\{\V\}$, AJIVE and SVD perform comparably, while SLIDE underperforms, likely due to its structural assumptions; SMILE attains the highest accuracy in every setting. For the accuracy of cluster means $\{\bmu\}$ and membership recovery, $k$-means generally outperforms hierarchical clustering when paired with the same embeddings, yet SMILE still clearly outperforms all competitors. As the number of embedding sources $L$ increases, all methods improve because additional views provide more information about the latent structure. As the vMF concentration $\kappa$ increases, performance again improves for all methods, most prominently for SMILE, since higher concentration yields more tightly clustered embeddings. In contrast, increasing the number of clusters $K$ makes the estimation and clustering more challenging, so all methods deteriorate, but SMILE remains the top performer. Notably, when $K = 100$, there are on average 10 codes per cluster, a scenario close to what we see in practice (e.g., the real data application in Section~\ref{sec:real_data}), and SMILE maintains a clear advantage, demonstrating its robustness in realistic high-resolution clustering settings.

Figure~S1 in Supplement~B reports the results for Setting 4. Because competing methods do not use relational pairs, their performance remains unchanged, whereas SMILE outperforms them even with the sparsest relational labels and exhibits improving estimation and clustering accuracy as more relational information becomes available, in line with the theoretical rates.

\section{Real Data Analysis}
\label{sec:real_data}

We evaluate SMILE using real-world EHR data from Mass General Brigham (MGB) and Veterans Affairs (VA), with the goal of harmonizing and grouping clinical codes through the integration of EHR-derived embeddings and language-based embeddings of those codes. Our evaluation focuses on two complementary objectives. First, we assess whether SMILE yields higher-quality latent embeddings than existing multi-view integration methods, using downstream tasks that measure recovery of known similarity/relatedness relationships and identification of disease-relevant features. Second, we examine whether SMILE improves clustering of synonymous or hierarchically related codes, particularly for aligning institution-specific codes to standardized ontologies. As in the simulations, we compare SMILE to SVD, AJIVE, and SLIDE for learning latent embeddings, and to each of these methods combined with $k$-means or hierarchical clustering for feature clustering.

We begin by describing the MGB and VA EHR datasets used in our analysis. The raw EHR codes were first standardized by mapping highly granular codes to higher-level code concepts, following the approach described in \cite{hong2021clinical}. From MGB, we include diagnoses mapped to PheCodes, medications mapped to RxNorm codes, and laboratory tests mapped to LOINC codes. From VA, we include the same three ontologies, along with two additional VA-specific laboratory codes, granular local lab codes (``OtherLab'') and grouped lab codes (``ShortName''). To increase cluster sizes and strengthen relational structure for embedding estimation, we further augment the dataset with additional PheCodes and RxNorm codes that share parents with existing MGB and VA codes. A summary of the resulting feature counts is provided in Table~\ref{tab:feature_counts}.

\begin{table}[ht]
\centering
\begin{tabular}{l|ccccc|c}
\hline
& PheCode & RxNorm & LOINC & OtherLab & ShortName & Total \\
\hline
MGB only   & 63      & 252    & 616   & 0        & 0    &931     \\
VA only    & 45      & 504    & 215   & 1814     & 94   &2672     \\
Overlap    & 1730    & 810    & 167   & 0        & 0   &2707      \\
Additionally added & 26   &  1112  & 0     & 0        & 0    &1138     \\
\hline
Total     & 1864    & 2678   & 998   & 1814     & 94   &  7448   \\
\hline
\end{tabular}
\caption{Summary of feature counts on the MGB and VA EHR datasets.}
\label{tab:feature_counts}
\end{table}

We incorporate $L=4$ embedding sources. Two are co-occurrence-based PMI-SVD embeddings ($1500$ dimensions each), computed separately from the MGB and VA EHR systems and therefore available only for codes observed at the corresponding institution. The remaining two are description-based embeddings derived from PLMs: SapBERT ($768$ dimensions; \citealt{liu2020self}) and BGE (1024 dimensions; \citealt{chen2024bge}), which are available for all codes. By combining EHR co-occurrence structure with semantic representations learned from biomedical text, we aim to jointly capture institution-specific usage patterns and cross-institution semantic consistency. 

Since PMI-SVD embeddings are missing for codes not observed at a given institution, competing methods that require complete embedding matrices are applied to imputed PMI-SVD representations. Specifically, we first apply SVD to concatenated SapBERT--BGE embeddings and then map the resulting representations into the PMI space via an orthogonal Procrustes transformation estimated using codes with observed PMI-SVD embeddings. In contrast, SMILE is trained directly on the original (non-imputed) embeddings, allowing it to accommodate heterogeneous, partially overlapping feature spaces without imputation.

To limit information leakage, we partition codes into training, validation, and test sets. For ontology-based codes (PheCode, RxNorm, and LOINC), the split is performed at the level of ontology parents so that all codes sharing a parent remain in the same subset. VA local laboratory codes (OtherLab and ShortName), which lack comparable hierarchies, are randomly assigned. The resulting split ratio is approximately 70/10/20, with feature counts summarized in Table~\ref{tab:feature_counts_by_split}. All codes enter SMILE through their source-specific embeddings.

\begin{table}[ht]
\centering
\begin{tabular}{l|ccccc|c}
\hline
& PheCode & RxNorm & LOINC & OtherLab & ShortName & Total \\
\hline
Training   & 1348      & 1795    & 703   &  1303       & 70    & 5219    \\
Validation    & 100      & 330    &  91  & 219     & 8   & 748    \\
Test    &  416   &  553   & 204   &  292       &  16  &  1481    \\
\hline
Total     & 1864    & 2678   & 998   & 1814     & 94   &  7448   \\
\hline
\end{tabular}
\caption{Summary of feature counts in the training, validation, and test sets.}
\label{tab:feature_counts_by_split}
\end{table}

For evaluation, we leverage the hierarchical structure of PheCode, RxNorm, and LOINC to define cluster targets. Two PheCodes are clustered together if they share the same integer parent; RxNorm codes are grouped by one-level-up Anatomical Therapeutic Chemical (ATC) class \citep{nahler2009anatomical}; and LOINC codes are grouped using first-level LOINC PART (LP) codes. VA-specific laboratory codes are mapped to these LP parents via curated OMOP relationships. The resulting clustering is highly granular: cluster sizes range from $1$ to $25$, with $41\%$ of codes in clusters of size at most $5$ and $68\%$ in clusters of size at most $10$. The number of clusters per category is selected by maximizing the Silhouette coefficient on $k$-means applied to SVD embeddings, yielding 600 clusters for PheCode, 570 for RxNorm, and 730 for laboratory codes, for a total of $K = 1900$. SMILE accepts a structured $n \times K$ prior probability matrix $\{\pi_{ik}\}$ that encodes partial cluster knowledge, in contrast to benchmark methods that cluster features separately within each category. For training codes, we treat their cluster memberships as known and set a hard prior with $\pi_{ik}=1$ on the true cluster and $0$ elsewhere. For validation and test codes, their true memberships are withheld from SMILE: validation memberships are reserved for hyperparameter tuning, and test memberships for final evaluation. For these codes, their priors are constructed by combining GPT-4--proposed candidate same-cluster pairs (Supplement~C) with spectral clustering under soft probabilistic assignments (Supplement~A). The resulting per-category probability matrices then stack into the block-structured prior matrix $\{\pi_{ik}\}$.

Similar pairs are constructed using ontology structure and mappings from the Unified Medical Language System (UMLS). With ontology hierarchies, two codes are labeled similar if they are first cousins, or if one code's parent is the other's grandparent. ``Hard negative'' pairs are defined if two codes are second cousins, or if one code's great-grandparent is the other's parent or grandparent. We further augment the positive set with established biomedical relationships from UMLS. To enrich the negative set, we sample pairs that are not ontology/UMLS positives but exhibit relatively high cosine similarity based on CODER embeddings. In total, approximately 6\% of code pairs are labeled as positive or negative similar pairs. Relatedness pairs are constructed similarly, incorporating UMLS and GPT-4-generated silver-standard labels (detailed procedure and prompt in Supplement~C); approximately 6\% of code pairs are labeled. All labeled pairs are then split into train, validation, and test sets. Training pairs are those with both codes in the training set; they comprise 58\% of all pairs and provide the only supervision SMILE receives. The remaining 42\% (with at least one held-out code) are randomly partitioned 25/75 into a validation set (10\% of all pairs, reserved for hyperparameter tuning) and a test set (32\%, reserved for final evaluation).

Since latent embedding quality and feature clustering quality reflect different levels of the model (feature- versus cluster-level) and admit no common optimum, we tune hyperparameters to balance both criteria rather than optimize either alone. Specifically, we evaluate each candidate configuration on the validation set using (i) the average clustering AMIs across PheCode, RxNorm, and laboratory categories on validation features, and (ii) the average areas under the ROC curves (AUCs) across similarity and relatedness on validation pairs. We select a configuration on the Pareto frontier of these two criteria, with selected latent dimension $r=60$. All methods are evaluated on the same test pairs and test features; since these test sets are not used during any method's training or hyperparameter selection, the comparison across methods is fair.

\subsection{Evaluating the Quality of Latent Embeddings}

We evaluate embedding quality through two downstream tasks: recovery of known similarity/relatedness relationships and selection of disease-relevant features. 

For the first task, we treat known similar/related pairs as positives and random pairs as negatives, and compute the AUC using embedding-based pairwise scores. For all benchmark methods, the score for a pair $(i,j)$ is the cosine similarity of the estimated feature-level embeddings $\wh{\V}_i$ and $\wh{\V}_j$, used for both similarity and relatedness AUC. For SMILE, the similarity score is likewise the cosine similarity of $\wh{\V}_i$ and $\wh{\V}_j$; the relatedness score is the cosine similarity of $\wh{\R}^{1/2}\wh{\V}_i$ and $\wh{\R}^{1/2}\wh{\V}_j$, reflecting the bilinear form in \eqref{eqn:relatedness_model}.

Results in Table~\ref{tab:similar_related_AUC} show that all multi-view methods achieve relatively strong similarity AUCs. The imputed institutional embeddings (MGB-PMI and VA-PMI) already perform competitively, which is unsurprising given that the imputation step itself fuses institutional co-occurrence information with PLM semantics via Procrustes alignment, thereby effectively combining two complementary information sources. SMILE matches or marginally exceeds the strongest baseline, and importantly does so without requiring an imputation step: it operates directly on the original partially overlapping vocabularies. This pattern reflects that ontology-defined similarity pairs are largely recoverable from well-integrated low-rank structure. However, the harder discriminative signal lies in the relatedness pairs. SMILE lifts relatedness AUC from 0.7762 to 0.9455 on the test set, a 22\% relative gain over SLIDE, the strongest baseline. Relatedness pairs span ontology branches and probe broader functional associations (e.g., between a disease and its treatment) that low-rank decomposition does not model directly, whereas SMILE's bilinear relatedness model captures these cross-cluster geometric relationships explicitly.

\begin{table}[ht]
\centering
\small
\resizebox{\textwidth}{!}{%
\begin{tabular}{ll|cccccccc}
\hline
 &          & MGB-PMI & VA-PMI & SapBERT & BGE    & SVD    & AJIVE  & SLIDE  & SMILE \\
\hline
\multirow{2}{*}{Sim}
 & Train & 0.8927 & 0.9106 & 0.7787 & 0.7841 & 0.8928 & 0.9140 & 0.9134 & \bf{0.9167} \\
 & Test  & 0.8562 & 0.8780 & 0.7340 & 0.7474 & 0.8583 & 0.8838 & 0.8828 & \bf{0.8878} \\
\hline
\multirow{2}{*}{Rel}
 & Train & 0.7628 & 0.7894 & 0.7506 & 0.6849 & 0.7664 & 0.7957 & 0.8016 & \bf{0.9459} \\
 & Test  & 0.7236 & 0.7670 & 0.7053 & 0.6606 & 0.7281 & 0.7686 & 0.7762 & \bf{0.9455} \\
\hline
\end{tabular}}
\caption{AUC comparison across embedding sources and methods. Train pairs are used for SMILE supervision, and test pairs are the 75\% of held-out pairs reserved for evaluation; the remaining 25\% of held-out pairs form the validation set used only for hyperparameter tuning and are not reported here. ``MGB-PMI'' and ``VA-PMI'' refer to imputed PMI-SVD embeddings derived from the MGB and VA EHR systems, respectively. ``Sim'' and ``Rel'' refer to similarity-based and relatedness-based labels, respectively.}
\label{tab:similar_related_AUC}
\end{table}

For the second task, we randomly select 11 PheCodes that appear in both the MGB and VA data as target diseases. For each disease, we construct a candidate feature set by taking the top 100 features across categories according to the aforementioned embeddings and augmenting them with 100 randomly sampled features as negative controls. For each embedding method, we compute the embedding cosine similarity between each candidate feature and the target disease as the feature's relevance score, and in parallel obtain GPT-4 semantic relevance scores based on textual feature descriptions. We then quantify agreement between embedding-based and GPT-4-based rankings using Spearman's rank correlation. Averaged over the 11 target diseases, the correlations are $\text{SVD} = 0.4283$, $\text{AJIVE} = 0.4110$, $\text{SLIDE} = 0.4064$, and $\text{SMILE} = 0.4407$, indicating that SMILE provides the most informative embeddings for identifying disease-relevant features.

\subsection{Clustering and Code Harmonization}

Clustering performance on the test set is measured by AMI, which is robust to many small, imbalanced clusters. Results for all methods are summarized in Table~\ref{tab:cluster_AMI}.

\begin{table}[ht]
\centering
\begin{tabular}{lccccccc}
\hline
& \multicolumn{3}{c}{$k$-means} & \multicolumn{3}{c}{hclust} &  \\
\cmidrule(lr){2-4}\cmidrule(lr){5-7}
           & SVD & AJIVE & SLIDE & SVD & AJIVE & SLIDE & SMILE \\
\midrule
PheCode      & 0.2337    & 0.2395      & 0.2379      & 0.2364    & 0.2372      & 0.2304      & \bf{0.3994}  \\
RxNorm       & 0.0801    & 0.0936      & 0.0860      & 0.0641    & 0.0765      & 0.0833      & \bf{0.1703}  \\
Lab          & 0.0685    & 0.0716      & 0.0718      & 0.0427    & 0.0623      & 0.0613      & \bf{0.1597}  \\
LOINC        & 0.0590    & 0.0669      & 0.0676      & 0.0394    & 0.0617      & 0.0606      & \bf{0.1135}  \\
VA local lab & 0.0559    & 0.0574      & 0.0582      & 0.0312    & 0.0521      & 0.0511      & \bf{0.1659}  \\
\hline
\end{tabular}
\caption{AMI of clustering performance on the test features for different methods. ``hclust'' refers to hierarchical clustering. ``Lab'' includes both LOINC codes and VA local laboratory codes (OtherLab and ShortName). ``VA local lab'' only includes VA local laboratory codes.}
\label{tab:cluster_AMI}
\end{table}

The ontology-based clustering problem is intrinsically difficult, with highly imbalanced cluster sizes and most codes in very small groups; in this setting, even a few misassignments create many pairwise disagreements and strongly depress chance-adjusted indices such as AMI, so even good methods rarely approach an AMI of 1. Across baselines, $k$-means and hierarchical clustering on SVD, AJIVE, and SLIDE yield broadly similar AMIs, indicating that existing multi-view methods offer limited benefit for recovering these fine-grained ontology clusters. In contrast, SMILE consistently outperforms all baselines: it improves AMI from 0.2395 to 0.3994 for PheCode (about 67\% gain), 0.0936 to 0.1703 for RxNorm (about 82\% gain), 0.0718 to 0.1597 for all lab codes (about 122\% gain), and 0.0676 to 0.1135 for LOINC (about 68\% gain). PheCode attains higher AMI because diagnosis clusters are coarser and more coherent than the more heterogeneous, fine-grained medication and lab clusters. Notably, for VA local laboratory codes, SMILE raises AMI from 0.0582 to 0.1659 (about 185\% gain), underscoring its utility for mapping institution-specific codes to standardized ontologies.

\section{Conclusion}
\label{sec:conclusion}

Multi-institutional studies face persistent interoperability challenges due to partially overlapping coding systems and heterogeneous data modalities. SMILE addresses this by learning a unified embedding space that integrates co-occurrence patterns, semantic embeddings, and biomedical knowledge. Rather than treating harmonization as a post-hoc mapping problem, SMILE aligns institution-specific codes directly within a shared latent representation, automatically harmonizing synonymous or related features across sites. This unified space enables patients to be represented using harmonized feature embeddings, mitigating coding discrepancies at the representation level and allowing downstream analyses to operate on semantically aligned patient profiles rather than institution-specific codes. 

Importantly, this framework is extensible beyond structured codes. Because SMILE operates at the embedding level, it can naturally incorporate narrative concepts extracted from clinical notes, such as symptom phrases, problem list entries, or contextual modifiers derived from large language models. Integrating structured codes and narrative-derived concepts within a unified embedding space would further reduce modality-specific fragmentation and move toward fully automated, representation-driven EHR harmonization. In this sense, SMILE provides not only improved clustering and embedding performance, but also a scalable infrastructure for cross-institutional, multi-modal patient representation in clinical and translational research.

{\small
\bibliographystyle{chicago}
\bibliography{align}

@article{balakrishnan2017statistical,
	author = {Balakrishnan, Sivaraman and Wainwright, Martin J and Yu, Bin},
	journal = {The Annals of Statistics},
	number = {1},
	pages = {77--120},
	publisher = {Institute of Mathematical Statistics},
	title = {Statistical guarantees for the {EM} algorithm: From population to sample-based analysis},
	volume = {45},
	year = {2017}}

@article{loh2015regularized,
	author = {Loh, Po-Ling and Wainwright, Martin J},
	journal = {Journal of Machine Learning Research},
	pages = {559--616},
	title = {Regularized {M}-estimators with nonconvexity: Statistical and algorithmic theory for local optima},
	volume = {16},
	year = {2015}}

@inproceedings{wang2015high,
	author = {Wang, Zhaoran and Gu, Quanquan and Ning, Yang and Liu, Han},
	booktitle = {Advances in Neural Information Processing Systems},
	pages = {2521--2529},
	title = {High dimensional {EM} algorithm: Statistical optimization and asymptotic normality},
	year = {2015}}

@article{ma2020implicit,
	author = {Ma, Cong and Wang, Kaizheng and Chi, Yuejie and Chen, Yuxin},
	journal = {Foundations of Computational Mathematics},
	number = {3},
	pages = {451--632},
	publisher = {Springer},
	title = {Implicit regularization in nonconvex statistical estimation: Gradient descent converges linearly for phase retrieval, matrix completion, and blind deconvolution},
	volume = {20},
	year = {2020}}

@article{rohe2011spectral,
	author = {Rohe, Karl and Chatterjee, Sourav and Yu, Bin},
	journal = {The Annals of Statistics},
	number = {4},
	pages = {1878--1915},
	title = {Spectral Clustering and the High-Dimensional Stochastic Blockmodel},
	volume = {39},
	year = {2011}}

@article{choi2012stochastic,
	author = {Choi, David S and Wolfe, Patrick J and Airoldi, Edoardo M},
	journal = {Biometrika},
	number = {2},
	pages = {273--284},
	publisher = {Oxford University Press},
	title = {Stochastic blockmodels with a growing number of classes},
	volume = {99},
	year = {2012}}

@article{day2017heterogeneous,
	author = {Day, Oscar and Khoshgoftaar, Taghi M.},
	journal = {Journal of Big Data},
	number = {1},
	pages = {29},
	title = {A survey on heterogeneous transfer learning},
	volume = {4},
	year = {2017}}

@article{feuz2015fsr,
	author = {Feuz, Kyle D. and Cook, Diane J.},
	journal = {ACM Transactions on Intelligent Systems and Technology},
	title = {Transfer Learning across Feature-Rich Heterogeneous Feature Spaces via Feature-Space Remapping (FSR)},
	year = {2015}}

@article{bao2023survey,
  title={A recent survey of heterogeneous transfer learning},
  author={Bao, Runxue and Sun, Yiming and Gao, Yuhe and Wang, Jindong and Yang, Qiang and Mao, Zhi-Hong and Ye, Ye},
  journal={arXiv preprint arXiv:2310.08459},
  year={2023}
}

@article{li2024multisource,
	author = {Li, Mengyan and Li, Xiaoou and Pan, Kevin and Geva, Alon and Yang, Doris and Sweet, Sara Morini and Bonzel, Clara-Lea and Ayakulangara Panickan, Vidul and Xiong, Xin and Mandl, Kenneth and others},
	journal = {NPJ Digital Medicine},
	number = {1},
	pages = {319},
	publisher = {Nature Publishing Group UK London},
	title = {Multisource representation learning for pediatric knowledge extraction from electronic health records},
	volume = {7},
	year = {2024}}

@article{brat2020international,
	author = {Brat, Gabriel A and Weber, Griffin M and Gehlenborg, Nils and Avillach, Paul and Palmer, Nathan P and Chiovato, Luca and Cimino, James and Waitman, Lemuel R and Omenn, Gilbert S and Malovini, Alberto and others},
	journal = {NPJ Digital Medicine},
	number = {1},
	pages = {109},
	publisher = {Nature Publishing Group UK London},
	title = {International electronic health record-derived COVID-19 clinical course profiles: the 4CE consortium},
	volume = {3},
	year = {2020}}

@article{li2023federated,
	author = {Li, Siqi and Liu, Pinyan and Nascimento, Gustavo G and Wang, Xinru and Leite, Fabio Renato Manzolli and Chakraborty, Bibhas and Hong, Chuan and Ning, Yilin and Xie, Feng and Teo, Zhen Ling and others},
	journal = {Journal of the American Medical Informatics Association},
	number = {12},
	pages = {2041--2049},
	publisher = {Oxford University Press},
	title = {Federated and distributed learning applications for electronic health records and structured medical data: a scoping review},
	volume = {30},
	year = {2023}}

@article{si2021deep,
	author = {Si, Yuqi and Du, Jingcheng and Li, Zhao and Jiang, Xiaoqian and Miller, Timothy and Wang, Fei and Zheng, W Jim and Roberts, Kirk},
	journal = {Journal of Biomedical Informatics},
	pages = {103671},
	publisher = {Elsevier},
	title = {Deep representation learning of patient data from Electronic Health Records (EHR): A systematic review},
	volume = {115},
	year = {2021}}

@article{garcia2025multi,
	author = {Garcia, Brittany and Hogarth, Michael and Wang, Yu and Zhu, Xi and Tu, Shin-Ping},
	journal = {Learning Health Systems},
	number = {4},
	pages = {e70039},
	publisher = {Wiley Online Library},
	title = {Multi-site research using electronic health record data: Lessons learned from a case study},
	volume = {9},
	year = {2025}}

@article{sittig2012survey,
	author = {Sittig, Dean F and Hazlehurst, Brian L and Brown, Jeffrey and Murphy, Shawn and Rosenman, Marc and Tarczy-Hornoch, Peter and Wilcox, Adam B},
	journal = {Medical Care},
	pages = {S49--S59},
	publisher = {LWW},
	title = {A survey of informatics platforms that enable distributed comparative effectiveness research using multi-institutional heterogenous clinical data},
	volume = {50},
	year = {2012}}

@article{bianchi2024all,
	author = {Bianchi, Diana W and Brennan, Patricia Flatley and Chiang, Michael F and Criswell, Lindsey A and D'Souza, Rena N and Gibbons, Gary H and Gilman, James K and Gordon, Joshua A and Green, Eric D and Gregurick, Susan and others},
	journal = {Nature Medicine},
	number = {2},
	pages = {330--333},
	publisher = {Nature Publishing Group US New York},
	title = {The All of Us Research Program is an opportunity to enhance the diversity of US biomedical research},
	volume = {30},
	year = {2024}}

@inproceedings{marwaha2024mobilizing,
	author = {Marwaha, Jayson S and Downing, Maren and Halamka, John and Abernethy, Amy and Franklin, Joseph B and Anderson, Brian and Kohane, Isaac and Wagholikar, Kavishwar and Brownstein, John and Haendel, Melissa and others},
	booktitle = {Healthcare},
	number = {2},
	organization = {Elsevier},
	pages = {100738},
	title = {Mobilizing data during a crisis: Building rapid evidence pipelines using multi-institutional real world data},
	volume = {12},
	year = {2024}}

@article{gan2025arch,
	author = {Gan, Ziming and Zhou, Doudou and Rush, Everett and Panickan, Vidul A and Ho, Yuk-Lam and Ostrouchovm, George and Xu, Zhiwei and Shen, Shuting and Xiong, Xin and Greco, Kimberly F and others},
	journal = {Journal of Biomedical Informatics},
	pages = {104761},
	publisher = {Elsevier},
	title = {Arch: Large-scale knowledge graph via aggregated narrative codified health records analysis},
	volume = {162},
	year = {2025}}

@inproceedings{choi2016multi,
	author = {Choi, Edward and Bahadori, Mohammad Taha and Searles, Elizabeth and Coffey, Catherine and Thompson, Michael and Bost, James and Tejedor-Sojo, Javier and Sun, Jimeng},
	booktitle = {Proceedings of the 22nd ACM SIGKDD International Conference on Knowledge Discovery and Data Mining},
	pages = {1495--1504},
	title = {Multi-layer representation learning for medical concepts},
	year = {2016}}

@article{arora2015latent,
	author = {Arora, Sanjeev and Li, Yuanzhi and Liang, Yingyu and Ma, Tengyu and Risteski, Andrej},
	journal = {arXiv preprint arXiv:1502.03520},
	title = {A latent variable model approach to PMI-based word embeddings},
	year = {2015}}

@inproceedings{pei2022graph,
	author = {Pei, Shichao and Yu, Lu and Yu, Guoxian and Zhang, Xiangliang},
	booktitle = {Proceedings of the ACM Web Conference},
	pages = {1104--1114},
	title = {Graph alignment with noisy supervision},
	year = {2022}}

@inproceedings{gopal2014mises,
	author = {Gopal, Siddharth and Yang, Yiming},
	booktitle = {International Conference on Machine Learning},
	organization = {PMLR},
	pages = {154--162},
	title = {Von mises-fisher clustering models},
	year = {2014}}

@inproceedings{barbaro2021sparse,
	author = {Barbaro, Florian and Rossi, Fabrice},
	booktitle = {2021 European Symposium on Artificial Neural Networks, Computational Intelligence and Machine Learning},
	title = {Sparse mixture of von Mises-Fisher distribution.},
	year = {2021}}

@article{zhou2022multiview,
	author = {Zhou, Doudou and Gan, Ziming and Shi, Xu and Patwari, Alina and Rush, Everett and Bonzel, Clara-Lea and Panickan, Vidul A and Hong, Chuan and Ho, Yuk-Lam and Cai, Tianrun and others},
	journal = {Journal of Biomedical Informatics},
	pages = {104147},
	publisher = {Elsevier},
	title = {Multiview incomplete knowledge graph integration with application to cross-institutional ehr data harmonization},
	volume = {133},
	year = {2022}}

@article{shi2021spherical,
	author = {Shi, Xu and Li, Xiaoou and Cai, Tianxi},
	journal = {Journal of the American Statistical Association},
	number = {536},
	pages = {1953--1964},
	publisher = {Taylor \& Francis},
	title = {Spherical regression under mismatch corruption with application to automated knowledge translation},
	volume = {116},
	year = {2021}}

@article{hong2021clinical,
	author = {Hong, Chuan and Rush, Everett and Liu, Molei and Zhou, Doudou and Sun, Jiehuan and Sonabend, Aaron and Castro, Victor M and Schubert, Petra and Panickan, Vidul A and Cai, Tianrun and others},
	journal = {NPJ Digital Medicine},
	number = {1},
	pages = {151},
	publisher = {Nature Publishing Group UK London},
	title = {Clinical knowledge extraction via sparse embedding regression (KESER) with multi-center large scale electronic health record data},
	volume = {4},
	year = {2021}}

@article{liu2020self,
	author = {Liu, Fangyu and Shareghi, Ehsan and Meng, Zaiqiao and Basaldella, Marco and Collier, Nigel},
	journal = {arXiv preprint arXiv:2010.11784},
	title = {Self-alignment pretraining for biomedical entity representations},
	year = {2020}}

@article{yuan2022coder,
	author = {Yuan, Zheng and Zhao, Zhengyun and Sun, Haixia and Li, Jiao and Wang, Fei and Yu, Sheng},
	journal = {Journal of Biomedical Informatics},
	pages = {103983},
	publisher = {Elsevier},
	title = {CODER: Knowledge-infused cross-lingual medical term embedding for term normalization},
	volume = {126},
	year = {2022}}

@article{alaux2018unsupervised,
	author = {Alaux, Jean and Grave, Edouard and Cuturi, Marco and Joulin, Armand},
	journal = {arXiv preprint arXiv:1811.01124},
	title = {Unsupervised hyperalignment for multilingual word embeddings},
	year = {2018}}

@inproceedings{grave2019unsupervised,
	author = {Grave, Edouard and Joulin, Armand and Berthet, Quentin},
	booktitle = {The 22nd International Conference on Artificial Intelligence and Statistics},
	organization = {PMLR},
	pages = {1880--1890},
	title = {Unsupervised alignment of embeddings with wasserstein procrustes},
	year = {2019}}

@article{lock2013joint,
	author = {Lock, Eric F and Hoadley, Katherine A and Marron, James Stephen and Nobel, Andrew B},
	journal = {The Annals of Applied Statistics},
	number = {1},
	pages = {523},
	title = {Joint and individual variation explained (JIVE) for integrated analysis of multiple data types},
	volume = {7},
	year = {2013}}

@article{feng2018angle,
	author = {Feng, Qing and Jiang, Meilei and Hannig, Jan and Marron, JS},
	journal = {Journal of Multivariate Analysis},
	pages = {241--265},
	publisher = {Elsevier},
	title = {Angle-based joint and individual variation explained},
	volume = {166},
	year = {2018}}

@article{zhou2015group,
	author = {Zhou, Guoxu and Cichocki, Andrzej and Zhang, Yu and Mandic, Danilo P},
	journal = {IEEE Transactions on Neural Networks and Learning Systems},
	number = {11},
	pages = {2426--2439},
	publisher = {IEEE},
	title = {Group component analysis for multiblock data: Common and individual feature extraction},
	volume = {27},
	year = {2015}}

@article{yang2016non,
	author = {Yang, Zi and Michailidis, George},
	journal = {Bioinformatics},
	number = {1},
	pages = {1--8},
	publisher = {Oxford University Press},
	title = {A non-negative matrix factorization method for detecting modules in heterogeneous omics multi-modal data},
	volume = {32},
	year = {2016}}

@article{gaynanova2019structural,
	author = {Gaynanova, Irina and Li, Gen},
	journal = {Biometrics},
	number = {4},
	pages = {1121--1132},
	publisher = {Oxford University Press},
	title = {Structural learning and integrative decomposition of multi-view data},
	volume = {75},
	year = {2019}}

@article{park2020integrative,
	author = {Park, Jun Young and Lock, Eric F},
	journal = {Biometrics},
	number = {1},
	pages = {61--74},
	publisher = {Oxford University Press},
	title = {Integrative factorization of bidimensionally linked matrices},
	volume = {76},
	year = {2020}}

@article{lock2022bidimensional,
	author = {Lock, Eric F and Park, Jun Young and Hoadley, Katherine A},
	journal = {The Annals of Applied Statistics},
	number = {1},
	pages = {193},
	title = {Bidimensional linked matrix factorization for pan-omics pan-cancer analysis},
	volume = {16},
	year = {2022}}

@article{yi2023hierarchical,
	author = {Yi, Sangyoon and Wong, Raymond Ka Wai and Gaynanova, Irina},
	journal = {Biometrics},
	number = {4},
	pages = {2933--2946},
	publisher = {Wiley Online Library},
	title = {Hierarchical nuclear norm penalization for multi-view data integration},
	volume = {79},
	year = {2023}}

@article{hu2024network,
	author = {Hu, Yaofang and Wang, Wanjie},
	journal = {Biometrika},
	number = {4},
	pages = {1221--1240},
	publisher = {Oxford University Press},
	title = {Network-adjusted covariates for community detection},
	volume = {111},
	year = {2024}}

@article{bramer1988international,
	author = {Br{\"a}mer, Gerlind R},
	journal = {World Health Statistics Quarterly. Rapport Trimestriel de Statistiques Sanitaires Mondiales},
	number = {1},
	pages = {32--36},
	title = {International statistical classification of diseases and related health problems. {Tenth revision.}},
	volume = {41},
	year = {1988}}

@article{mcdonald2003loinc,
	author = {McDonald, Clement J and Huff, Stanley M and Suico, Jeffrey G and Hill, Gilbert and Leavelle, Dennis and Aller, Raymond and Forrey, Arden and Mercer, Kathy and DeMoor, Georges and Hook, John and others},
	journal = {Clinical Chemistry},
	number = {4},
	pages = {624--633},
	publisher = {Oxford University Press},
	title = {LOINC, a universal standard for identifying laboratory observations: a 5-year update},
	volume = {49},
	year = {2003}}

@article{liu2005rxnorm,
	author = {Liu, Simon and Ma, Wei and Moore, Robin and Ganesan, Vikraman and Nelson, Stuart},
	journal = {IT Professional},
	number = {5},
	pages = {17--23},
	publisher = {IEEE},
	title = {RxNorm: prescription for electronic drug information exchange},
	volume = {7},
	year = {2005}}

@article{chen2024bge,
	author = {Chen, Jianlv and Xiao, Shitao and Zhang, Peitian and Luo, Kun and Lian, Defu and Liu, Zheng},
	journal = {arXiv preprint arXiv:2402.03216},
	number = {5},
	title = {Bge m3-embedding: Multi-lingual, multi-functionality, multi-granularity text embeddings through self-knowledge distillation},
	volume = {4},
	year = {2024}}

@article{tsybakov2004optimal,
	author = {Tsybakov, Alexander B},
	journal = {The Annals of Statistics},
	number = {1},
	pages = {135--166},
	publisher = {Institute of Mathematical Statistics},
	title = {Optimal aggregation of classifiers in statistical learning},
	volume = {32},
	year = {2004}}

@inproceedings{fromont2006functional,
	author = {Fromont, Magalie and Tuleau, Christine},
	booktitle = {International Conference on Computational Learning Theory},
	organization = {Springer},
	pages = {94--108},
	title = {Functional classification with margin conditions},
	year = {2006}}

@incollection{nahler2009anatomical,
	author = {Nahler, Gerhard},
	booktitle = {Dictionary of Pharmaceutical Medicine},
	pages = {8--8},
	publisher = {Springer},
	title = {Anatomical therapeutic chemical classification system (ATC)},
	year = {2009}}

@article{zhou2025representation,
  title={Representation learning to advance multi-institutional studies with electronic health record data from {US} and {France}},
  author={Zhou, Doudou and Tong, Han and Wang, Linshanshan  and others},
  journal={Nature Communications},
  year={2026},
  publisher={Nature Publishing Group UK London}
}

@article{holland1983stochastic,
	author = {Holland, Paul W and Laskey, Kathryn Blackmond and Leinhardt, Samuel},
	journal = {Social Networks},
	number = {2},
	pages = {109--137},
	publisher = {Elsevier},
	title = {Stochastic blockmodels: First steps},
	volume = {5},
	year = {1983}}

@article{lei2015consistency,
	author = {Lei, Jing and Rinaldo, Alessandro},
	journal = {The Annals of Statistics},
	pages = {215--237},
	publisher = {JSTOR},
	title = {Consistency of spectral clustering in stochastic block models},
	year = {2015}}
}

\clearpage
\end{document}